\documentclass[letter,twocolumn]{jpsj2} 
\def\co{Ba(Fe$_{0.9}$Co$_{0.1}$)$_{2}$As$_{2}$}
\def\cox{Ba(Fe$_{1-x}$Co$_{x}$)$_{2}$As$_{2}$}
\def\bak{Ba$_{1-x}$K$_{x}$Fe$_{2}$As$_{2}$}
\def\ba{BaFe$_{2}$As$_{2}$}
\def\afeas{$A$Fe$_{2}$As$_{2}$}

\def\ro{$R$FeAsO$_{1-x}$F$_{x}$}

\title{Possible superconductivity above 25 K in single crystalline Co-doped BaFe$_{2}$As$_{2}$}

\author{Yasuyuki \textsc{Nakajima}$^{1,2}$\footnote[1]{E-mail : yasuyuki@ap.t.u-tokyo.ac.jp}, Toshihiro \textsc{Taen}$^{1}$, and Tsuyoshi \textsc{Tamegai}$^{1,2}$}

\inst{$^{1}$Department of Applied Physiscs, The University of Tokyo, 7-3-1 Hongo, Bunkyo-ku, Tokyo 113-8656, Japan\\
$^{2}$JST, Transformative Research-Project on Iron Pnictides (TRIP), 7-3-1 Hongo, Bunkyo-ku, Tokyo 113-8656, Japan
}

\abst{We present superconducting properties of single crystalline {\co} by measuring magnetization, resistivity, upper critical field, Hall coefficient, and magneto-optical images. The magnetization measurements reveal fish-tail hysteresis loop at high temperatures and relatively high critical current density above $J_{c}=10^{5}$ A/cm$^{2}$ at low temperatures. Upper critical field determined by resistive transition is anisotropic with anisotropic parameter $\sim$ 3.5. Hall effect measurements indicate that {\co} is a multiband system and the mobility of electron is dominant. The magneto-optical imaging reveals prominent Bean-like penetration of vortices although there is a slight inhomogeneity in a sample. Moreover, we find a distinct superconductivity above 25 K, which leads us to speculate that higher transition temperature can be realized by fine tuning Co-doping level. }

\kword{Iron arsenide, Magneto-optical imaging, Critical current density, Hall coefficient, Upper critical field}

\begin{document}

\maketitle


Since the discovery of high-$T_{c}$ iron-based oxypnictide superconductor LaFeAsO$_{1-x}$F$_{x}$ with $T_{c}\sim$ 26 K \cite{kamih08}, other iron-based superconductors are sought for to obtain higher transition temperature. In rare-earth substituted iron-oxypnictides {\ro} ($R$ = rare earth), the transition temperature has been increased up to 55 K \cite{ren08}. In these iron oxypnictides, superconductivity occurs by introducing electrons in the (FeAs)$^{-}$ layers by substituting F for O.
Following these discoveries, oxygen-free iron-arsenide {\afeas} ($A=$ Ba, Sr, Ca) is discovered. These materials show superconductivity by substituting alkali metal, such as Na, K and Cs, for $A$ resulting in the introduction of holes in the (FeAs)$^{-}$ layers \cite{rotte08a, chen08,sasma08, wu08}.  In hole-doped oxygen-free iron-arsenides, transition temperature is raised up to $\sim$ 38 K \cite{rotte08a, chen08}. On the other hand, very recent studies reveal that the electron-doping by substitution of Co or Ni, which have one or two excess $d$ electrons compare to Fe in conducting layers, induces superconductivity in oxygen-free iron-arsenide \cite{sefat08,li08}. In fact, NMR measurements reveal that Co atoms donate electrons without creating localized moments \cite{ning08a}. Although the highest transition temperature in electron-doped {\ba} is reported to be $\sim$ 23 K \cite{ni08}, which is lower than that in hole-doped {\ba}, the fact that the substitution of transition metals leads to occurrence of superconductivity contrasts strongly with a drastic suppression of $T_{c}$ in cuprates \cite{fujis89}. However, the detailed study of the transition temperature as a function of Co-doping is limited \cite{ni08}. It is an open question whether the highest $T_{c}$ in electron-doped {\ba} is increased further or not.

In this paper, we have prepared the single crystalline sample of {\co} and present its superconducting properties by studying the magnetization, resistivity, upper critical field, Hall coefficient, and magneto-optical images. We address the possibility of further enhancement of transition temperature by fine-tuning Co-doping level.


\begin{figure}[b]
\begin{center}
\includegraphics[width=8cm]{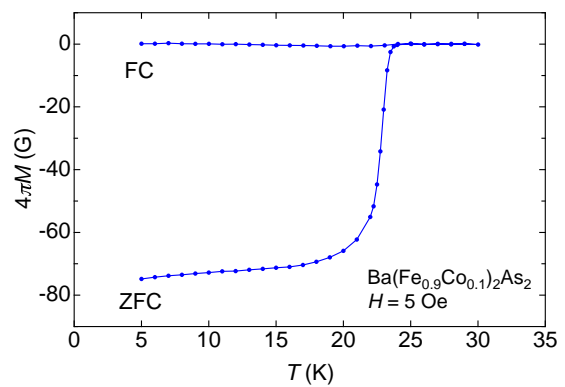}
\caption{Temperature dependence of the zero-field-cooled (ZFC) and field-cooled (FC) magnetization at $H=$ 5 Oe in {\co}.  }
\label{FIG1}
\end{center}
\end{figure}

Single crystalline samples with nominal composition {\co} were grown by FeAs/CoAs self-flux method. FeAs and CoAs were prepared by placing mixtures of As pieces and Fe/Co powder in a silica tube and reacted at 1065 $^{\circ}$C for 10 hours after heating at 700 $^{\circ}$C for 6 hours. A mixture with the ratio Ba : FeAs : CoAs = 1 : 4.5 : 0.5 was  placed in alumina crucible with quartz fiber as a cup. The whole assembly sealed in a large silica tube, and heated up to 1150$^{\circ}$C for 10 hours followed by slow-cooling down to 1090$^{\circ}$C at the rate of $\sim$ 1.3 $^{\circ}$C/h, after which the silica tube was put into a centrifuge to separate crystals from flux. The typical size of resulting crystals is $\sim$2$\times$2$\times$0.05~mm$^{3}$. Magnetization was measured by a commercial SQUID magnetometer (MPMS-XL5, Quantum Design). The resistivity measurements were performed by four-contact method and Hall coefficient were measured by the standard six-wire configuration. The Hall voltage was obtained from the antisymmetric part of the transverse voltage by subtracting the positive and negative magnetic field data. Magneto-optical images were obtained by using the local-field-dependent Faraday effect in the in-plane magnetized garnet indicator film and employing a differential method \cite{soibe00,yasug02}.


Figure 1 shows the temperature dependence of zero-field-cooled (ZFC) and field-cooled (FC) magnetization at 5 Oe. Very sharp transition starting at $T_{c}\sim$ 24 K is observed. It should be noted that  this value of $T_{c}$ is the highest among those reported to date \cite{sefat08,ni08,ning08a,yamam08,prozo08,gordo08,ning08}.

\begin{figure}[t]
\begin{center}
\includegraphics[width=8cm]{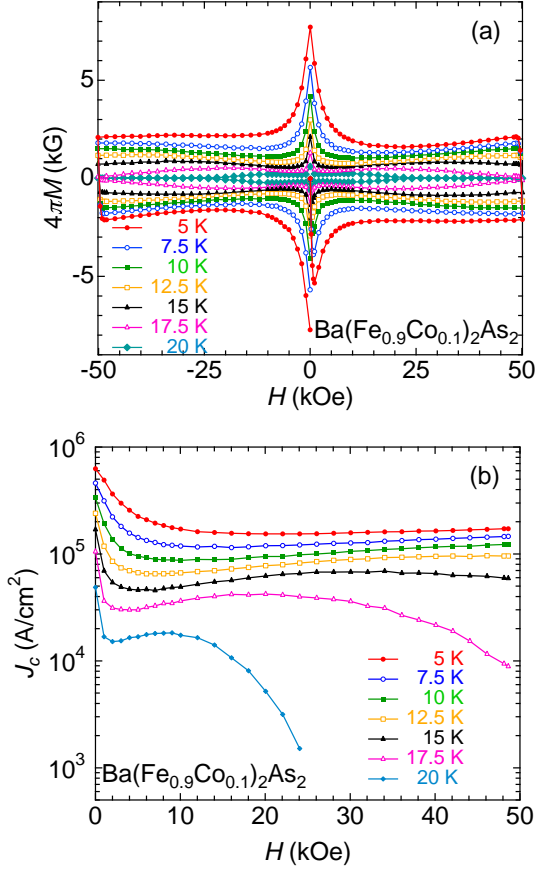}
\caption{(a) Field dependence of magnetization in {\co} at 5, 7.5, 10, 12.5, 15, 17.5, and 20 K. (b) Field dependence of critical current density obtained by the data shown in Fig. 2. (a) in {\co} at 5, 7.5, 10, 12.5, 15, 17.5,  and 20 K. }
\label{FIG2}
\end{center}
\end{figure}

Figure 2(a) depicts the magnetization at several temperatures as a function of field. At high temperatures above 15 K, fish-tail magnetization is observed, which is very similar to YBa$_{2}$Cu$_{3}$O$_{7-\delta}$ single crystals \cite{daeum90}. From the magnetization hysteresis loop, we can obtain the critical current density $J_{c}$ using the Bean model with the assumption of field-independent $J_{c}$. According to the Bean model, $J_{c}$ is given by
\begin{equation}
	J_{c}= 20 \frac{\Delta M}{a(1-a/3b)},
\end{equation}
where $\Delta M$ is $M_{down}-M_{up}$, $M_{up}$ and $M_{down}$ are the magnetization when sweeping fields up and down, respectively, $a$ and $b$ are sample widths with $a<b$. Figure 2(b) shows the field dependence of $J_{c}$ obtained by the data shown in Fig. 2(a) using Eq. (1) and effective sample dimension with $a\sim$ 0.55 mm and $b\sim$ 0.63 mm. At low temperatures, $J_{c}$ is larger than 10$^{5}$ A/cm$^{2}$. This value is much larger than that reported in Ref. \citen{sefat08} and about twice as large as those reported in Refs. \citen{yamam08} and \citen{prozo08}. While the present value of $J_{c}$ is about one order of magnitude smaller than the typical value for YBa$_{2}$Cu$_{3}$O$_{7-\delta}$  single crystals \cite{tameg92}, it is well within the range for practical applications.

\begin{figure}[t]
\begin{center}
\includegraphics[width=8cm]{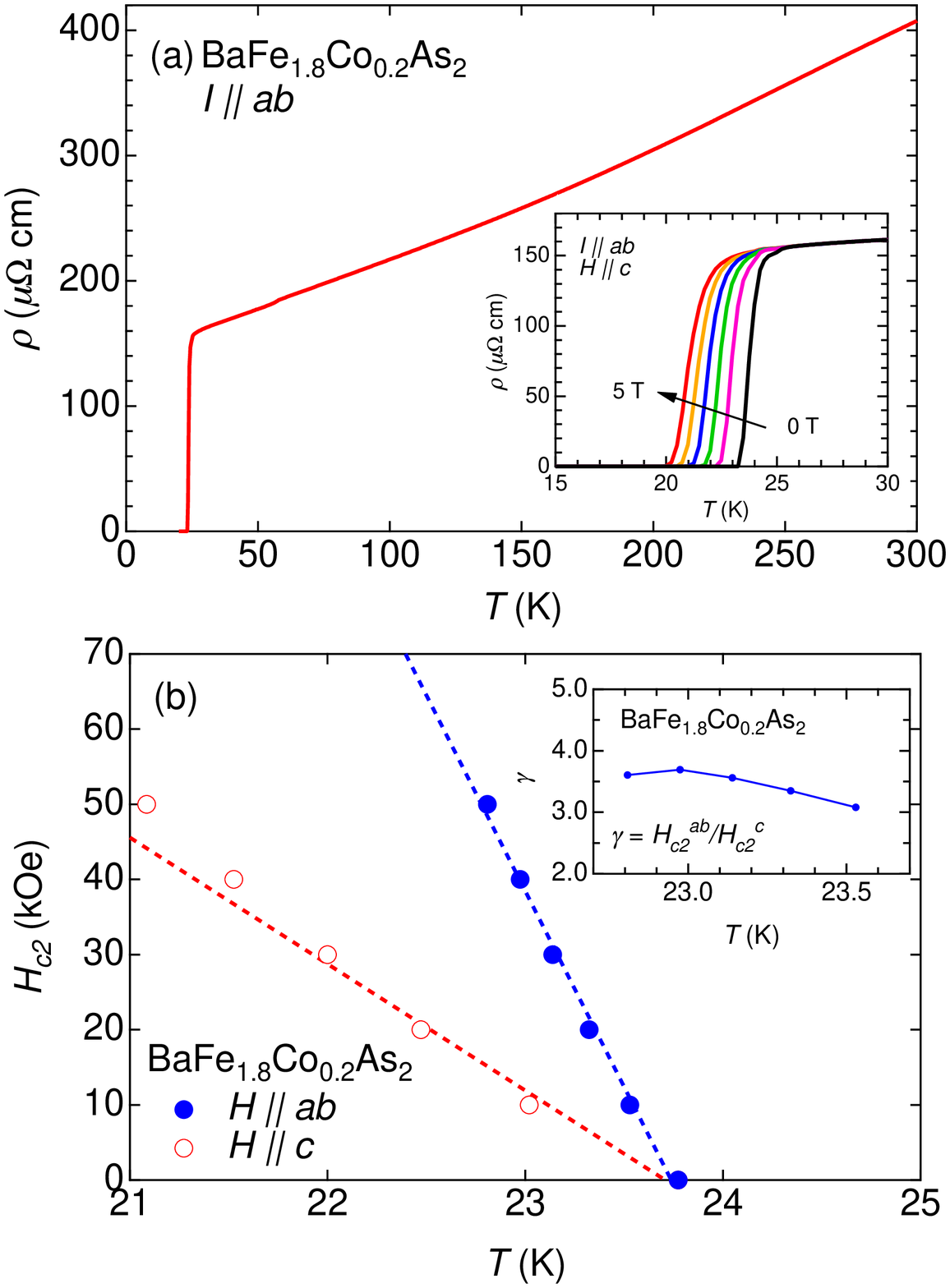}
\caption{(a) Temperature dependence of the zero-field resistivity in {\co}. Inset: low-temperature in-plane resistivity data at $H=$ 0, 10, 20, 30, 40, and 50 kOe along $c-$axis. (b)  Temperature dependence of upper critical field along $ab$-plane ($\bullet$) and $c$-axis ($\circ$) obtained by the midpoint of  resistive transition in {\co}. Dashed lines are linear fits to the data. Inset: the temperature dependence of  anisotropy of the upper critical field along $ab-$ and $c-$directions $\gamma\equiv H_{c2}^{ab}/H_{c2}^{c}$.}
\label{FIG3}
\end{center}
\end{figure}

Figure 3 (a) shows temperature dependence of in-plane zero-field resistivity in {\co}.  With decreasing temperature from 300 K, the resistivity decreases monotonically and then drops suddenly at $T_{c}$. There is no anomaly accompanied by a spin-density-wave transition reported in the parent material {\ba} \cite{rotte08}, which indicates that the transition is suppressed by Co-doping. The residual resistivity ratio $\rho(300~\mathrm{K})/\rho(T_{c})$ is $\sim$ 2.6, which is comparative to that reported before \cite{sefat08,ni08, prozo08,ning08}. Inset of Fig.3(a) depicts low-temperature in-plane resistivity data at $H=$ 0, 10, 20, 30, 40, and 50 kOe along $c-$axis. With increasing fields, $T_{c}$ decrease and the transition width is broadened only slightly. We plot the upper critical field $H_{c2}$ along $ab-$ and $c-$ directions determined by the midpoint of resistive transition as a function of temperature in Fig. 3(b). The values of slope of $H_{c2}$ along $ab-$ and $c-$direction at $T_{c}$ are -52.1 and -16.8 kOe/K, respectively. From the Werthamer-Helfand-Hohenberg theory \cite{werth66}, which describes orbital depairing field of conventional dirty type-II superconductors, we can obtain the values of $H_{c2}(0)=0.69T_{c}|dH_{c2}/dT|_{T=T_{c}}$ $\sim 850$ and $\sim 280$ kOe along $ab-$ and $c-$direction, respectively. These values are very high and comparative to hole-doped iron-arsenide superconductors {\bak} \cite{altar08}. The inset of Fig. 3(b) shows the anisotropy of the upper critical field $\gamma\equiv H_{c2}^{ab}/H_{c2}^{c}$. The value of $\gamma$ is $\sim$ 3-4 and, slightly increases with decreasing temperature. This value is similar to hole-doped iron arsenide \cite{ni08a}.
\begin{figure}[t]
\begin{center}
\includegraphics[width=8cm]{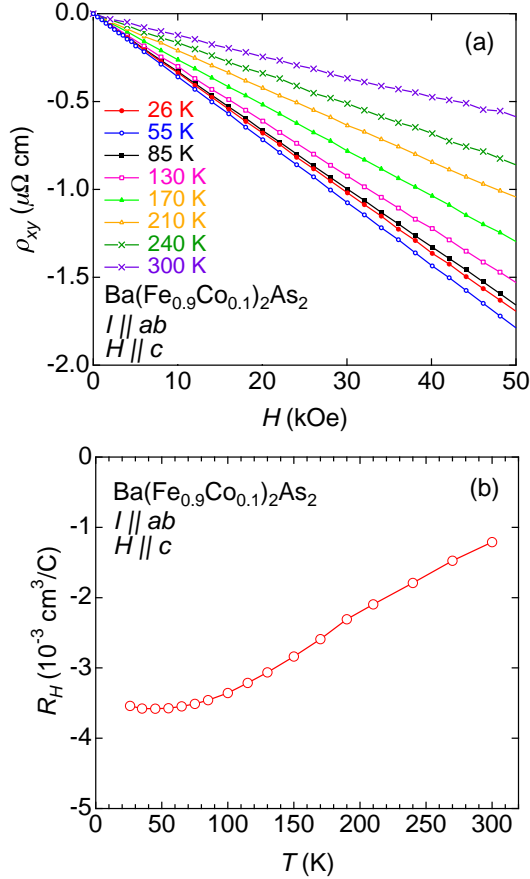}
\caption{(a) The Hall resistivity $\rho_{xy}$ as a function of field at several temperatures. (b) Temperature dependence of the Hall coefficient in {\co}. }
\label{FIG4}
\end{center}
\end{figure}

Hall effect measurements strongly support that electron is introduced by Co-doping and {\co} is multiband system. Figure 4(a) shows the Hall resistivity $\rho_{xy}$ at several temperatures as a function of field. In the present temperature region, $\rho_{xy}$ is negative and shows $H-$ linear dependence. We plot the Hall coefficient $R_{H}$ in {\co} obtained from $\rho_{xy}$ as a function of temperature in Fig. 4(b). The sign of $R_{H}$ is  negative in the whole temperature range while that in the hole-doped {\ba} is positive \cite{wu08a}, which is consistent with introduction of electrons by Co-substitution for Fe. The values of $R_{H}$ decrease with decreasing temperature from 300 K to $\sim$ 60 K and slightly increase after showing a broad minimum at $\sim$ 50 K. The absolute value of $R_{H}$ just above $T_{c}$ is three times larger than that at 300 K.  In a simple single band system, the Hall coefficient is written as, $R_{H}=1/nq$, where $q$ is a charge of a carrier and $n$ is a carrier density. $R_{H}$ is almost $T$-independent. By contrast, the Hall coefficient in multiband system, for instance, consisting of electron and hole bands, is given by, $R_{H}= (n_{h}\mu_{h}^{2}-n_{e}\mu_{e}^{2})/(e(n_{h}\mu_{h}+n_{e}\mu_{e})^{2})$, where $n_{h}$ ($n_{e}$) is a density of holes (electrons) and $\mu_{h}$ ($\mu_{e}$) is a mobility of holes (electrons). The Hall coefficient in multiband system can be temperature dependent. The obtained results indicate that {\co} is multi-band system similar to the parent material {\ba} \cite{liu08} and mobility of the electron produced by Co-doping becomes dominant with decreasing temperature.

\begin{figure}[t]
\begin{center}
\includegraphics[width=8cm]{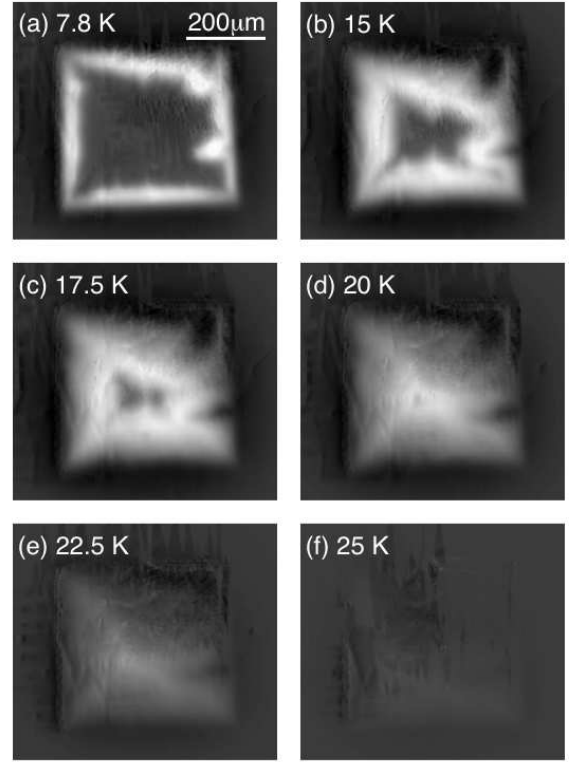}
\caption{Magneto-optical imaging of remanent state after applying $H=$ 690 Oe in {\co} at (a) 7.8 K, (b) 15 K, (c) 17.5 K, (d) 20 K, (e) 22.5 K, and (f) 25 K. }
\label{FIG5}
\end{center}
\end{figure}

Figure 5 shows magneto-optical images of {\co} in the remanent state at several temperatures after cycling the field up to 690 Oe for five seconds. While the sample for the magneto-optical experiments is different (0.58$\times$0.45$\times$0.03~mm$^{3}$) from that for the magnetization measurements, both crystals show similar properties. The bright region in figures corresponds to the area trapping the vortices. Dark faint zig-zag features originate from in-plane domains of the garnet film and have nothing to do with the flux distribution. At lower temperatures, vortices penetrated from edges of the crystal cannot reach the center due to large shielding current. However, prominent penetrations of vortices are observed near the defects close to the right edge of the sample. The penetrations of vortices depelop with increasing temperature. It should be noted that the image at 25 K apparently shows existence of superconductivity in the lower part of the sample. In fact, the zero-field resistivity starts to drop at $\sim$ 25 K as shown in the inset of Fig. 3(a). The highest  transition temperature in {\cox} reported so far is $\sim 23$ K \cite{ni08}. The difference can originate from the difference in Co concentration and/or its distribution. It may be possible to obtain even higher transition temperature in this system by fine-tuning the Co-doping level.

In summary, we have presented a systematic study of magnetization, resistivity, upper critical field, Hall coefficient, and magneto-optical imaging in single crystalline {\co}. The magnetization measurements reveal fish-tail hysteresis loop at high temperatures and the relatively high critical current density larger than $10^{5}$ A/cm$^{2}$. Upper critical field obtained by resistive transition is anisotropic. Its anisotropy is about 3.5. Hall effect measurements indicate that {\co} is multiband system and the mobility of electron is dominant. The magneto-optical imaging reveals prominent Bean-like penetrations of vortices although there is slight inhomogeneity in a sample.  We also find a distinct superconductivity above 25 K, which leads us to expect that higher transition temperature can be realized by fine-tuning Co-doping.

\end{document}